\documentclass[sigconf]{acmart}

\usepackage{booktabs} 
\usepackage[skip=0pt]{caption} 

\setcopyright{rightsretained}

\acmConference[WSDM 2018]{WSDM 2018}{February 5--9, 2018}{Los Angeles, USA} 
\acmYear{2018}
\copyrightyear{2018}

\acmPrice{15.00}

\usepackage[subtle, wordspacing=normal, tracking=normal, bibnotes, charwidths=normal, leading, indent, lists, paragraphs=normal, mathspacing=normal, bibliography=normal]{savetrees}
\usepackage{flushend}

\usepackage{multirow}
\usepackage{enumitem}

\begin{document}

\title{Neural Networks for Information Retrieval}
\titlenote{Slides and other materials will be made available online at \url{http://nn4ir.com}.}

\author{Tom Kenter}
\authornote{Corresponding author.}
\affiliation{%
  \institution{Booking.com}
  \city{Amsterdam, The Netherlands}
}
\email{tom.kenter@gmail.com}

\author{Alexey Borisov}
\affiliation{%
  \institution{Yandex}
  \city{Moscow, Russia}
}
\email{alborisov@yandex-team.ru}

\author{Christophe Van Gysel}
\affiliation{%
  \institution{University of Amsterdam}
  \city{Amsterdam, The Netherlands}
}
\email{cvangysel@uva.nl}

\author{Mostafa Dehghani}
\affiliation{%
  \institution{University of Amsterdam}
  \city{Amsterdam, The Netherlands}
}
\email{dehghani@uva.nl}

\author{Maarten de Rijke}
\affiliation{%
  \institution{University of Amsterdam}
  \city{Amsterdam, The Netherlands}
}
\email{derijke@uva.nl}

\author{Bhaskar Mitra}
\affiliation{%
  \institution{Microsoft, University College London}
  \city{Cambridge, UK}
}
\email{bmitra@microsoft.com}

\renewcommand{\shortauthors}{T. Kenter et al.}

\settopmatter{printacmref=true}
\fancyhead{}

\copyrightyear{2018} 
\acmYear{2018} 
\setcopyright{rightsretained} 
\acmConference[WSDM 2018]{WSDM 2018: The Eleventh ACM International Conference on Web Search and Data Mining }{February 5--9, 2018}{Marina Del Rey, CA, USA}
\acmBooktitle{WSDM 2018: WSDM 2018: The Eleventh ACM International Conference on Web Search and Data Mining , February 5--9, 2018, Marina Del Rey, CA, USA}
\acmDOI{10.1145/3159652.3162009}
\acmISBN{978-1-4503-5581-0/18/02}

\maketitle


\subsection*{Abstract}
Machine learning plays a role in many aspects of modern IR systems, and deep learning is applied in all of them.
The fast pace of modern-day research has given rise to many approaches to many IR problems.
The amount of information available can be overwhelming both for junior students and for experienced researchers looking for new research topics and directions.
The aim of this full-day tutorial is to give a clear overview of current tried-and-trusted neural methods in IR and how they benefit IR.


\subsection*{Motivation}

Prompted by advances of deep learning in computer vision, neural networks (NNs) have resurfaced as a popular machine learning paradigm in many other directions of research, including IR.
Recent years have seen NNs being applied to all key parts of the typical modern IR pipeline, such as 
click models, 
core ranking algorithms, 
dialogue systems,
entity retrieval, 
knowledge graphs, 
language modeling, 
question answering, 
and
text similarity. 

An advantage that sets NNs apart from many learning strategies employed earlier, is their ability to work from raw input data.
Where designing features used to be a crucial aspect of newly proposed IR approaches, the focus has shifted to designing network architectures instead.
Hence, many architectures and paradigms have been proposed, such as auto-encoders, recursive networks, recurrent networks, convolutional networks, various embedding methods, and deep reinforcement learning.
This tutorial aims to provide an overview of the main network architectures currently applied in IR and to show how they relate to previous work.

We expect the tutorial to be useful both for academic and industrial researchers and practitioners who either want to develop new neural models, use them in their own research in other areas or apply the models described here to improve actual IR systems.


\subsection*{Brief outline of the topics to be covered}
Table~\ref{table:schedule} gives an overview of the time schedule of the tutorial.
The total time is 6 hours, plus breaks.
\begin{table}[h]
  \centering
  \caption{Time schedule for NN4IR tutorial}
  \label{table:schedule}
  \begin{tabular}{@{}l l l l@{}}
    \toprule
    \multicolumn{2}{c}{Morning} & \multicolumn{2}{c}{Afternoon} \\
    \cmidrule(r){1-2}\cmidrule(l){3-4}
    Preliminaries          & 45 min. &  Recommender systems    & 45 min. \\
    Semantic matching      & 45 min. &     Modeling user behavior & 45 min. \\
    Learning to rank       & 45 min. &     Generating responses   & 45 min. \\
    Entities               & 45 min.  &   Industry insights      & 45 min. \\
   \bottomrule
  \end{tabular}
\end{table}
We bring a team of six lecturers, all with their specific areas of specialization.
Each session will have two expert lecturers (indicated by their initials below) who will together present the session.

{
\setlength{\baselineskip}{3pt}


\paragraph{\textbf{Preliminaries} [TK, MdR]}
The recent surge of interest in deep learning has given rise to a myriad of model architectures.
Different though the inner structures of NNs can be, many building blocks are shared.
In this preliminary session, we focus on key concepts, all of which will be referred to multiple times in subsequent sessions.
In particular we will cover distributed representations/embeddings \cite{mikolov2013efficient}, fully-connect layers, convolutional layers \cite{krizhevsky2012imagenet}, recurrent networks \cite{mikolov2010recurrent} and sequence-to-sequence models \cite{sutskever2014sequence}.


\paragraph{\textbf{Semantic matching} [CVG, BM]}
The problem of matching based on textual descriptions arises in many retrieval systems. The traditional IR approach involves computing lexical term overlap between query and document \citep{robertson1995okapi}. However, a vocabulary gap occurs when query and documents use different terms to describe the same concepts \citep{Li2014semantic}.
Semantic matching methods bridge the vocabulary gap by matching concepts rather than exact word occurrences.
We cover neural network-based methods that learn to provide a semantic matching signal supervised fashion \citep{lu2013deep,Mitra:2016:www,huang2013learning,mitra2016desm}, semi-supervised \citep{Dehghani:2017, dehghani2017avoiding}, and unsupervised \citep{vulic2015monolingual,kenter2015short,Zuccon2015nntm,Ganguly2015generalizedlm,le2014distributed,VanGysel2017nvsm,VanGysel:2016:cikm,VanGysel:2016:www,Ai2016doc2veclm,kenter2016siamesecbow}.


\paragraph{\textbf{Learning to rank} [AB, MD]}
Capturing the notion of relevance for ranking needs to account for different aspects of the query, the document, and their relationship.
Neural methods for ranking can use manually crafted query and document features, and combine them with regards to a ranking objective.
Moreover, latent representations of the query and document can be learned in situ. 
We cover scenarios with different levels of supervision---unsuper\-vised \cite{VanGysel:2016:www, VanGysel:2016:cikm, salakhutdinov2009semantic}, semi/weakly-supervised \cite{Dehghani:2017,dehghani2017avoiding, szummer2011semi}, or fully-supervised using labeled data \cite{Mitra:2016:www} or interaction data \cite{huang2013learning}.


\paragraph{\textbf{Entities} [CVG, TK]}
Entities play a central role in modern IR systems \citep{Dietz2016tutorial}.
We cover neural approaches to solving the basic task of named entity recognition~\citep{Collobert2008unified,Chiu2015ner,Lample2016ner}, as well learning representations in an end-to-end neural model for learning a specific task like entity ranking for expert finding~\citep{VanGysel:2016:www}, product search~\citep{VanGysel:2016:cikm} or email attachment retrieval \citep{VanGysel2017proactive}.
Furthermore, work related to knowledge graphs will be covered, such as graph embeddings \citep{bordes2011learning,Wang2014knowledge,Zhao2015}.


\paragraph{\textbf{Recommender systems} [MdR, BM]}
Deep learning has also found its way into recommender systems.
We cover learning of item (products, users) embeddings~\citep{barkan2016item2vec,grbovic2015e-commerce,vasile2016meta-prod2vec}, as well as deep collaborative filtering using different deep learning techniques and architectures~\citep{wang2015collaborative,cheng2016wide}.
Furthermore, NN-based feature extraction from content (such as images, music, text)~\citep{bansal2016ask,oord2013deep,mcauley2015image-based}, and session-based recommendations with RNNs~\citep{Dehghani:CIKM2017, hidashi2016session-based,quadrana2017personalizing} will be covered.

\paragraph{\textbf{Modeling user behavior} [AB, MdR]}
Modeling user browsing behavior plays an important role in modern IR systems.
Accurately interpreting user clicks is difficult due to various types of bias.
Many click models based on Probabilistic Graphical Models (PGMs) have been proposed~\cite{chuklin2015click}.
Recently, it was shown that recurrent neural networks can learn to account for biases in user clicks directly from click-through, i.e., without the need for a predefined set of rules as is customary for PGM-based click models~\cite{Borisov:2016:click}.


\paragraph{\textbf{Generating responses} [TK, MD]}

Recent inventions such as smart home devices, voice search, and virtual assistants provide new ways of accessing information.
They require a different response format than the classic ten blue links.
Examples are conversational and dialog systems \citep{li2016deep,vinyals2015neural} or machine reading and question answering tasks, where the model either infers the answer from unstructured text \citep{Hermann:2015:nips, weston2015towards,hewlett2016wikireading,amn2017kenter,serban2016generating,kenter2017byte_level}  or generates natural language given structured data, like data from knowledge graphs or from external memories~\citep{Graves:2014:neural, Ahn:2017,Lebret:2016, mei:2015}.


\paragraph{\textbf{Industry insights} [AB, BM]}
Where the focus of academic papers can be on a specific subtask, industry approaches have to ensure that a system works from start to end.
As a result, extra challenges are involved concerning the user experience.
For example in Google's SmartReply system \cite{kannan2016smart_reply} the neural model at the core of the system is embedded in a much larger framework of non-neural methods to make sure quality and efficiency requirements are met.

}

\bibliographystyle{abbrvnatnourl}
\bibliography{wsdm-tutorial-nn4ir} 

\end{document}